\documentclass[11pt,a4paper]{article}
\usepackage{amsmath,amssymb}
\usepackage{graphicx}
\begin{document}

\title{Computer algebra calculations in supersymmetric electrodynamics}

\author{I.E.Shirokov $\vphantom{\Big(}$
\medskip\\
{\small{\em Moscow State University,}}\\
{\small{\em Faculty of Physics, Department of Theoretical Physics,}}\\
{\small{\em 119991, Moscow, Russia.}}
}

\maketitle

\begin{abstract}
We propose a new symbolic algorithm and a C++ program for generating and calculating supersymmetric Feynman diagrams for ${\cal N}=1$ supersymmetric electrodynamics regularized by higher derivatives in four dimensions. According to standard rules, the program generates all diagrams that are necessary to calculate a specific contribution to the two-point Green function of matter superfields in the needed order, and then reduces the answer to the sum of Euclidean momentum integrals. At the moment, the program was used to calculate the anomalous dimension in ${\cal N}=1$ supersymmetric quantum electrodynamics, regularized by higher derivatives, in the three-loop approximation.
\end{abstract}

\section{Introduction}
\hspace*{\parindent}

Attempts to create software for calculations within the framework of quantum perturbation theory have been made for more than fifty years (see, for example, \cite{Campbell:1970tm}, as well as a review \cite{Gerdt:1978pv}). Of course, the early programs were very limited, mostly, they considered quantum electrodynamics. The main problem for the development of such programs was the extremely low performance of machines of the day. However, since the 90s, the rapid development of this kind of software has begun. Some programs created at that time have not lost relevance until now (for example, \cite{Nogueira:1991ex, Kublbeck:1990xc}).

Various programs for calculations in high-energy physics can be divided into several groups. Firstly, there are programs designed to calculate various processes, usually within the framework of the Standard Model. They are usually limited to the tree (PHEGAS\cite{Papadopoulos:2000tt}, O'mega\cite{Moretti:2001zz}, madevent\cite{Maltoni:2002qb}, FDC\cite{Wang:2004du}, CompHEP/CalcHEP\cite{CompHEP:2004qpa, Belyaev:2012qa}, WHIZARD\cite{Kilian:2007gr}, Herwig++\cite{Bahr:2008pv}, SHERPA\cite{Gleisberg:2008ta}) or maximally to the one-loop approximation (GoSam\cite{Cullen:2014yla}, aMC@NLO\cite{Alwall:2014hca}). Some of these programs are also the event generators \cite{Papadopoulos:2000tt, Maltoni:2002qb, Bahr:2008pv, Gleisberg:2008ta}, which can simulate the result of an experiment. These programs are good because they mainly generate diagrams, amplitudes and even calculate Feyman integrals. However, they are very limited by possible theories, the number of loops (usually no more than one), as well as the field of application (mostly these are scattering cross sections, decay widths, etc.).

The second group of programs is related to the generation of amplitudes in a more general case. One of the most famous programs is QGRAPH\cite{Nogueira:1991ex}. This program has been known for about 30 years, but is still actively used in various studies. It generates all diagrams in any order of the perturbation theory, outputs them in some symbolic form, and also counts combinatorial coefficients. At the same time, it neither generates amplitudes nor draws diagrams. Another well-known program of this kind is FeynArts\cite{Kublbeck:1990xc, Hahn:2000kx}. It generates diagrams and amplitudes for a given theory. It also depicts them graphically using a special package for \LaTeX. As a disadvantage, it can be pointed that it is limited to the three loop approximation.

We should also mention various computer algebra systems. Analytical calculations of diagrams have their own specifics because in addition to simple algebraic operations (which can be performed using well-known systems such as Mathematica\cite{Math}, Maple\cite{Maple}, Schoonschip\cite{Veltman:1991xb}, as well as FORM\cite{Ruijl:2017dtg}), it is necessary to analyze tensor structures, calculate traces of gamma matrices, etc. There is a number of packages and programs for this. For example GiNaC\cite{Vollinga:2005pk}, Cadabra\cite{Peeters:2006kp}, Redberry\cite{Bolotin:2013qgr}, FeynCalc\cite{Shtabovenko:2020gxv}.

There are some programs that calculate Feynman momentum integrals in the D-dimention, which are formed after construction of  the amplitudes. Mostly, the following approach to their calculation is implemented. Firstly, tensor integrals are reduced to scalar integrals by standard methods, and then using such techniques as integration by parts in D-dimention \cite{Chetyrkin:1981qh} and the application of Lorentz invariance scalar integrals are reduced to a small number of typical master integrals. This process is implemented in the programs AIR\cite{Anastasiou:2004vj}, FIRE \cite{Smirnov:2019qkx}, LiteRed\cite{Lee:2013mka}, Reduze\cite{Studerus:2009ye} and Kira\cite{Maierhofer:2017gsa}. Also there are well-known programs for calculating master integrals, such as AMBRE\cite{Dubovyk:2016ocz}, FIESTA\cite{Smirnov:2015mct} and SecDec\cite{Borowka:2015mxa}. We should also mention the package for Schoonschip Mincer \cite{Gorishnii:1989gt}, which was firstly used to calculate the four-loop beta function using dimensional regularization \cite{Gorishnii:1990kd}.

Finally, there are software packages that combine the generation of diagrams and amplitudes, operations with them and taking integrals. For example aITALC\cite{Lorca:2004dk}, FeynMaster\cite{Fontes:2019wqh}, HepLib\cite{Feng:2021kha}, \texttt{tapir} \cite{Gerlach:2022qnc}. Basically, all these packages are based on the generation of diagrams by the QGRAPH program, and then on the analysis of the results by other programs mentioned above.

Some of the mentioned programs, among other things, are adapted to work with supersymmetric theories \cite{Hahn:2001rv, Wang:2004du, CompHEP:2004qpa, Gleisberg:2008ta}. All of them work within the framework of the Minimal Supersymmetric Standard Model in terms of component fields. However, two known programs for working with superfields in the superspace exist. They are SUSYCAL\cite{Kreuzberger:1989qq} program written using PASCAL language, as well as a package for Mathematica SusyMath\cite{Ferrari:2007sc}. These programs can work with superspace expressions, that are generated by supergraphs. In theory, they can simplify them to momentum integrals, but these projects are not being developed and are currently unavailable for download.

Thus, we can note that despite the significant progress in this area, there is a noticeable lack of software for working within supersymmetric theories in terms of superfields. Even existing programs need integration with those that generate diagrams. In addition, the generation of graphs in the supersymmetric case has its own specifics. To this end, the author has developed computer-algebraic approaches for working with superfields in the superspace. Using them, the author created a program, that is capable to generate Feynman diagrams in terms of superspace, as well as to perform various operations with them, after which the result is output in the form of standard Feynman integrals. At the moment, the program can be used for calculating the anomalous dimension of the matter superfields in ${\cal N}=1$ supersymmetric quantum electrodynamics regularized by higher covariant derivatives in four dimensions.

\section{${\cal N}=1$ superspace formalism}
\hspace*{\parindent}\label{Section_RGFs_With_Multiple_Couplings}

${\cal N}=1$ superspace is a space with the coordinates $(ct,x,y,z,\theta)$, where $\theta$ is a Majorana spinor. Spinors in four dimensions are transformed according to a special spinor law with respect to the Lorentz group\footnote{You can read more about spinors, for example, in \cite{ClassField}}. Due to using of the superspace supersymmetry is explicit, even at the quantum level. The spinor indices are raised and lowered using charge conjugation matrices:
\begin{equation}
\theta^a \equiv \theta_b C^{ba};\qquad \theta_a = \theta^b C_{ab}.
\end{equation}
The supersymmetric covariant derivative is usually introduced as follows:
\begin{equation}
\bar{D}_{\dot{a}} = \frac{\partial}{\partial\bar\theta^{\dot{a}}} - i(\gamma^\mu)_{\dot{a}}{}^b\theta_b\, \partial_\mu .
\end{equation}
In this case, the spinor indices are indicated by Latin letters, the right ones without a dot, and the left ones with a dot, also the left spinors are marked by a bar. For example, $\bar{D}_{\dot{a}} $ is the left supersymmetric covariant derivative, and the right one is denoted as $D_{a}$. At the same time , it is true for squares of derivatives:
\begin{equation}
D^2=D^{a}D_{a} \quad \bar D^2 =\bar D^{\dot a} \bar D_{\dot a}.
\end{equation}
The usual fields in this approach are components of superfields. So, the gauge field is a component of the real superfield $V(x^{\mu},\theta)$, spinor, and scalar superfields are components of chiral or antichiral fields ($\phi(x^{\mu},\theta)$ and $\phi^{*}(x^{\mu},\theta)$, respectively), which, by definition, satisfy the conditions:
\begin{equation}
\bar D_{\dot a} \phi=0, \quad D_{a} \phi^{*}=0.
\end{equation}
In addition, when we construct supersymmetric actions, integration is introduced with respect to $\theta$ variables. In our notation , it can be defined as follows:
\begin{equation}\label{int}
\int d^2 \bar \theta= \frac{1}{2}\bar D^2 , \quad \int d^2 \theta= -\frac{1}{2} D^2 \nonumber
\end{equation}
\begin{equation}
 \int d^4 \theta=\int d^2 \bar \theta d^2 \theta .
\end{equation}

\section{Perturbation theory in supersymmetric theories}
\hspace*{\parindent}\label{Section_MSSM_NSVZ}

We now consider how the standard perturbation theory works in the superspace\footnote{You can read about the usual perturbation theory in QFT, for example, in \cite{Bogolyubov:1959bfo}}. First of all, it is necessary to consider how the action of the theory is written in terms of superfields. There are 2 invariants with respect to supersymmetry transformations, which can be written as follows:
\begin{equation}
S_{1}=\int d^4 x \, d^4\theta \, \mathbb{V}, \quad S_{2}=\int d^4 x \, d^2\theta \,\Phi +\mbox{c.c.}
\end{equation}
where $\mathbb{V}$ is a real superfield and $\Phi$ is a chiral superfield.

It is most convenient to carry out quantization by the functional integral method. The main element of this approach, from which various quantities can be obtained, is the generating functional $Z$, which is constructed as follows:
\begin{equation}\label{gf1}
Z=\int {\cal D}(\mbox{superfields})\, e^{iS+iS_{\operatorname{sources}}},
\end{equation}
where ${\cal D}(\mbox{superfields})$ is a measure of the functional integration.
The action can be represented as:
\begin{equation}
S=S^{(2)}+S_{int}(\phi,V),
\end{equation}
where $S^{(2)}$ is a contribution to the action quadratic in fields, and $S_{int}$ is a sum of contributions of degree higher than 2. Next, terms with sources for real and chiral fields look like this:
\begin{equation}
S_{\operatorname{sources}}=j\phi+JV
\end{equation}
In this case (\ref{gf1}) can be rewritten as follows:
\begin{equation}\label{gf}
Z=e^{i S_{int}(\frac{1}{i}\frac{\delta}{\delta j},\frac{1}{i}\frac{\delta}{\delta J})} \int {\cal D}(\mbox{superfields})\, e^{i(S^{(2)}+S_{\operatorname{sources}})}
\end{equation}
The remaining Gaussian integral is taken using standard methods. The interaction term series expansion is interpreted graphically using Feynman diagrams.

After the expansion, one can obtain structures which generate propagators in diagrams. For example, for gauge superfield sources:
\begin{equation}\label{prop}
\frac{\delta}{\delta J_{1}}\frac{\delta}{\delta J_{2}}Z_{0}(J)
\end{equation}
At the same time , the following structures arise:
\begin{equation}
\frac{\delta J_{1}}{\delta J_{2}}=\delta^{8}_{12}=\delta^4(x_1-x_2)\delta^4_{12}(\theta)
\end{equation}
where $\delta^4_{12}(\theta)$ is defined as 
\begin{equation}
\delta^4_{12}(\theta)=\delta^4(\theta_1-\theta_2)=4(\theta_1-\theta_2)^2(\bar{\theta}_1-\bar{\theta}_2)^2.
\end{equation}
With this definition, standard rules for a superspatial delta-function are obtained:
\begin{equation}\label{deltaint}
\int d^4 \theta_{1}\delta^4_{12}(\theta)=1; \quad \int d^4 \theta_{1} f(\theta_1) \delta^4_{12}(\theta)=f(\theta_2).
\end{equation}
The following relation holds for the sources of chiral fields:
\begin{equation}
\frac{\delta j_{1}}{\delta j_{2}}=-\frac{D^2}{2}\delta^{8}_{12}.
\end{equation}
In addition, $D$-derivatives can exist in the vertices and in the generating functional itself.

Sometimes, it is necessary to consider the effective action. It is defined as follows. First, the generating functional of  the connected Green's functions is constructed:
\begin{equation}
W[\mbox{Sources}]=-i\ln Z[\mbox{Sources}],
\end{equation}
and the effective action $\Gamma[\mbox{superfields}] $ is obtained from $W[\mbox{Sources}] $ by Legendre transformation:
\begin{equation}\label{effact}
\Gamma=W-S_{\operatorname{sources}} \vert_{\mbox{sources} \longrightarrow\mbox{superfields}}, \nonumber
\end{equation}
\begin{equation}
\frac{\delta W}{\delta j_{i}}=\varphi_{i}.
\end{equation}
In fact, the generating functional $W$ removes all disconnected diagrams from $Z$, and $\Gamma$ removes one-particle reducible diagrams. These are diagrams that can be divided by cutting a single internal line. As an example, in Fig. \ref{ris:image10} the upper diagram is one-particle irreducible, and the lower one is one-particle reducible.
\section{Supersymmetric ``D-algebra''}
It can be seen that a specific expression for a Feynman diagram will contain some superspatial delta-functions $\delta^{8}_{12}$, to which supersymmetric covariant derivatives act, while the entire expression is integrated over the superspace. A typical expression has the form
\begin{equation}\label{dexpmpl}
\int d^8 x_1  ...d^8 x_n \, D\bar{D}...D\delta^8_{12}\times...\times D\bar{D}...D\delta^8_{nm}.
\end{equation}
At the same time, in accordance with \cite{West:1990tg}, the following algorithm for working with this expression can be constructed:

\newpage
\begin{figure}[h]
\centering
\includegraphics[scale=1]{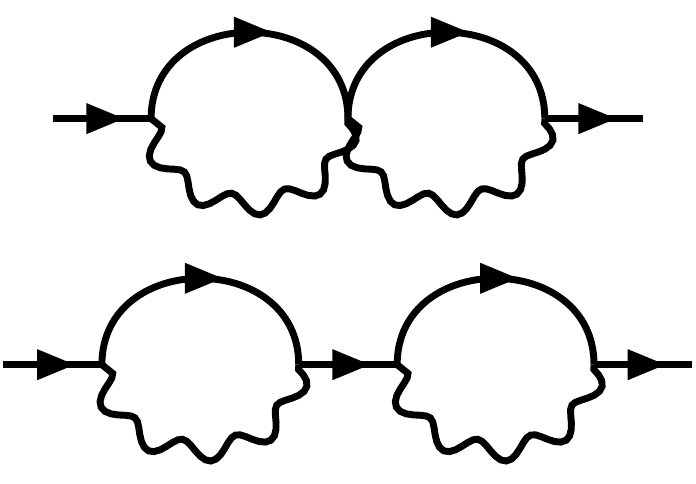}
\caption{One-particle (ir)reducible diagrams}
\label{ris:image10}
\end{figure}
\vspace{1em}
\begin{enumerate}
\item
We make `integration by parts". One can transfer the external $D$-derivative from any element to another one corresponding to the same point. Thus, it is possible to clear a superspatial delta-function from derivatives and integrate it according to (\ref{deltaint}).
\item
It is also possible to rearrange derivatives among themselves in accordance with the rules of supersymmetric `D-algebra" in four dimensions:
\begin{equation}\label{r1}
\lbrace D_{a}, \bar{D}_{\dot{b}}\rbrace= 2i(\gamma^{\mu})_{a\dot{b}}\partial_{\mu},
\end{equation}
\begin{equation}
\lbrace D_{a}, D_{b}\rbrace= 0,
\end{equation}
\begin{equation}
\lbrace \bar{D}_{\dot{a}}, \bar{D}_{\dot{b}}\rbrace= 0,
\end{equation}
\begin{equation}
[ D_{a}, \bar{D}^2]= 4i(\gamma^{\mu})_{a}^{\,\,\dot{b}}\bar{D}_{\dot{b}}\partial_{\mu},
\end{equation}
\begin{equation}
[ D_{a}, D^2]= 0,
\end{equation}
\begin{equation}
[ \bar{D}^2, D^2]= 4i(\gamma^{\mu})^{\dot{b}a}[D_{a},\bar{D}_{\dot{b}}]\partial_{\mu},
\end{equation}
\begin{equation}
\bar{D}^2 D^2 \bar{D}^2=-16\partial^2 \bar{D}^2,
\end{equation}
\begin{equation}\label{r2}
D^2 \bar{D}^2 D^2 =-16\partial^2 D^2.
\end{equation}
These identities help to reduce a number of the derivatives.
\item
If there are several identical superspatial delta-functions in the expression, then the following identities should be applied:
\begin{equation}\label{r3}
\delta^8_{12} D^{\alpha} \delta^8_{12}=0, \quad\mbox{if}\,\,\alpha\leq 3
\end{equation}
\begin{equation}\label{r4}
\delta^8_{12}D^2 \bar{D}^2 \delta^8_{12}=4 \delta^8_{12} \delta^4(x_1-x_2)
\end{equation}
\end{enumerate}
As a result of these steps, it is possible to remove most of the integrations and the superspacial delta-functions, while covariant derivatives acting on external superfields may remain. Also there will be a certain number of traces of the gamma matrices, which are evaluated using the standard proceedures. After that, the expressions under consideration take the form of ordinary momentum integrals.
\section{The algorithm implemented by the program}
\hspace*{\parindent}
Let us briefly describe the algorithm of the program. We calculate the contribution to the effective action (\ref{effact}). To do this, one can use Eq. (\ref{gf}) if one excludes from it all one-particle reducible and disconnected diagrams and also makes substitution of external fields according to Legendre transformation (\ref{effact}). In accordance with the above, the program operates as follows.
\begin{enumerate}
\item
In Eq. (\ref{gf}), the exponent with the interaction vertices is expanded to the desired order of the perturbation theory. At the same time, terms where external superfields are substituted in all possible ways are added.
\item
Further, in all the resulting contributions we try to pair derivatives with respect to sources, for example as in the formula (\ref{prop}). If this fails in the end, then the option is discarded, if it works, a new expression is created in which derivatives with respect to sources are grouped in pairs.
\item
We check the resulting graphs for one-particle reducibility. To do this, we use a function that removes each of the propagators in turn and then checks the connectivity. After each check we either remove the graph or leave it.
\item
At this stage we switch to the momentum representation, generate a list of loop momenta based on the expnsion order. Loop momenta are basis momenta, from which other momenta of internal lines can be expressed. Each loop momenta is encoded by a prime number. Next, we assign a momentum to each propagator, based on the conservation law for all vertices of a diagram, and replace the paired derivatives sources by a specific expression for the propagator, which contains an operator acting on a superspacial delta-function (it can also contain momenta and constants). After that, each graph is also checked for compliance with the conservation law in each point. If something is incorrect, the program throws an error.
\item
Now the expressions are ready to work with the $D$-algebra. Each expression is processed separately. We consider each operator and choose the one with the fewest covariant derivatives, then we begin to transfer all the operators from it using integration by parts, until we completely clear the superspacial delta-function on which it acts. In this case, a lot of new expressions are obtained, and then we separately process each of them. When the superspacial delta-function is cleared, we look for the same one and, if found, apply the rules (\ref{r1})---(\ref{r2}) to reduce the number of derivatives, and then Eqs. (\ref{r3}) and (\ref{r4}) to eliminate the superspacial delta-function. If there is no such superspacial delta-function, the integration is simply removed. We act in this way until we eliminate all superspacial delta-functions. Then, if the external momenta are not equal to zero, we do the same for matter superfields, taking into account their chirality.
\item
Now we have expressions with external fields, momenta, and other structures, as well as gamma matrices, delta-symbols, and charge conjugation matrices with spinor indices. At this step, we contract all indices from the delta symbols and charge conjugation matrices, and then calculate the traces of the gamma matrices. In general, in our notation, this is not a completely pure trace of gamma matrices, it also contains $\gamma_{5}$, which can lead to structures containing the antisymmetric tensor $\epsilon_{\mu\nu\lambda \delta}$, but it will always be contracted with loop momenta and, therefore, will not contribute up to the five-point Green's functions, which we do not consider yet \footnote{More details about the algebra of $\gamma$-matrices can be found, for example, in \cite{ClassField}}. It should be noted that unlike calculations in D-dimension \cite{Tarasov:1980nsq}, we work in four dimensions and therefore all standard relations for $\gamma$-matrices are valid in our case.
\item
After that, we get contributions in the form of squares and scalar products of momenta. We represent a momentum not by a string, but by a number. This is done for the convenience of working with it, because one can use standard algorithms for finding the greatest common divisor, etc. Each momentum is given by a prime number, and the sum of the momenta corresponds to their product, so each number uniquely sets the sum. For example: we have momenta $k$, $l$ and $q$. Let us assign $k$-number 2, $l$-number 3, $q$-number 5. Then, for example, $k+l$ will be 6, and $k+l+q$ will be 30. Mostly, when constructing diagrams, all lines can be assigned exactly the sum of impulses. If at some point this is not the case, then, for example, a square can be represented by $(k-l)^2=-(k+l)^2+2k^2+2l^2$. In this case, it is easier to work with full squares, separating them from the general expression. As an example, the following expression can be considered:
\begin{equation}\label{im}
(q+k+l)^{\mu} (q+k)_{\mu}=(q+k)^2+l^{\mu} (q+k)_{\mu}\nonumber
\end{equation}
\begin{equation}
=(q+k)^2+\frac{1}{2}(q+k+l)^2-\frac{1}{2}l^2-\frac{1}{2}(q+ k)^2
\end{equation}
In the notation of the program, these are the products of momentum 30 ($q+k+l$) and 10 ($q+k$). First we look for the greatest common divisor, it is equal to 10 ($q+k$), and then multiply the quotient of each with GCD (in this case it is 3($l$) and 1(0)) with each of the other factors (10 ($q+k$) and 30 ($q+k+l$), respectively). If the greatest common divisor is 1, then we decompose into perfect squares in the standard way (as in the second part (\ref{im})). At the end, for each individual graph, we try to collect terms.
\item
At the last stage, we try to collect terms between different graphs. In this case, we also use some possible changes of variables, since all final expressions are assumed to be integrated over loop momenta. We implement substitutions of the following type:
\begin{equation}
k \rightarrow k+l, \quad k \rightarrow k+l+q \quad \mbox{etc.}
\end{equation}
as well as all possible permutations of momenta:
\begin{equation}
k \leftrightarrow l, \quad k \leftrightarrow q \quad \mbox{etc.}
\end{equation}
In doing so, we act as follows. We try to make all possible replacements of these two classes, and for each replacement we put in correspondence a certain weight (which represents the sum of the values of all momenta). Then we choose such a replacement, which has the smallest weight. In the general case, this minimum is not unique, but in most situations it is. Then we carry out a direct comparison of all expressions and collect terms. This greatly simplifies the expression.
\end{enumerate}

A few general remarks should also be made. At almost all stages, we collect terms as they are represented at that time. This is usually done by direct comparison of the expressions term by term. At the first stage, where the pairing of derivatives with respect to sources takes place, the reduction of similar ones is carried out by permuting points and further comparison until all permutations have been enumerated. This speeds up the program by orders of magnitude. Also, at the stage of generation and direct processing of each graph, the program is divided into threads using the OpenMP library \cite{OpenMp}.

\section{Examples}
\hspace*{\parindent}\label{Section_MSSM_NSVZ22}

Let us describe examples of using the program. We will consider ${\cal N}=1$ supersymmetric electrodynamics \cite{West:1990tg, Gates:1983nr, Buchbinder:1998qv} with $N_{f}$ types of matter superfields (we will further call them flavors) regularized by higher derivatives \cite{Slavnov:1971aw,Slavnov:1972sq,Krivoshchekov:1978xg, West:1985jx}. Regularization is carried out in four dimensions. In the massless limit the action of the original theory has the form:
\begin{equation}
\begin{aligned}
&S=\frac{1}{4e_{0}^2}\mbox{Re}\int d^4x d^2\theta W^{a}W_{a}+\frac{1}{4} \sum_{\alpha=1}^{N_{f}} \int d^{4} x d^{4} \theta\left(\phi_{\alpha}^{*} e^{2 V} \phi_{\alpha}+\widetilde{\phi}_{\alpha}^{*} e^{-2 V} \widetilde{\phi}_{\alpha}\right)
\end{aligned}
\end{equation}
where $e_{0}$ is a bare coupling constant, $V$ is a gauge superfield and $\phi_{\alpha}\,$ are $N_{f}$ chiral matter superfields.
In order to regularize the theory, it is necessary to add to the action the regularization function $R(\frac{\partial^{2}}{\Lambda^{2}})$, which satisfy the following conditions
\begin{equation}\label{cond}
R(0)=1 , \quad
R(\infty)=\infty.
\end{equation}
Then the regularized action of the theory takes the form
\begin{equation}\label{act}
\begin{aligned}
&S_{\operatorname{reg}}=\frac{1}{4e_{0}^2}\mbox{Re}\int d^4x d^2\theta W^{a}R(\frac{\partial^{2}}{\Lambda^{2}})W_{a}+\frac{1}{4} \sum_{\alpha=1}^{N_{f}} \int d^{4} x d^{4} \theta\left(\phi_{\alpha}^{*} e^{2 V} \phi_{\alpha}+\widetilde{\phi}_{\alpha}^{*} e^{-2 V} \widetilde{\phi}_{\alpha}\right).
\end{aligned}
\end{equation}
It is also necessary to add a gauge fixing term:
\begin{equation}
S_{\mathrm{gf}}=-\frac{1}{32 e_{0}^{2} \xi_{0}} \int d^{4} x d^{4} \theta D^{2} V K\left(\partial^{2} / \Lambda^{2}\right) \bar{D}^{2} V,
\end{equation}
here function $K(\frac{\partial^{2}}{\Lambda^{2}})$ satisfy the same conditions (\ref{cond}) as function $R(\frac{\partial^{2}}{\Lambda^{2}})$. For regularization in the one-loop approximation, one must also add the Pauli-Villars determinants \cite{Slavnov:1977zf}. They can be represented as path integrals over the corresponding chiral superfields $\Phi$ and $\tilde{\Phi}$:
\begin{equation}
(\det(PV,M))^{-1}=\int D\Phi D\tilde{\Phi} \, \exp(i S_{PV}) ,
\end{equation}
where the action for Pauli-Villars superfields is given by the expression
\begin{equation}
\begin{aligned}
&S_{PV}=-\frac{1}{4} \int d^4x d^4\theta (\Phi^{*} e^{2V} \Phi + \tilde{\Phi}^{*} e ^{-2V} \tilde{\Phi})+\frac{1}{2} \int d^4x d^2\theta (M \Phi \tilde{\Phi} + c.c.) ,
\end{aligned}
\end{equation}
and $M=a\Lambda$ is the mass of the Pauli-Villars superfields.

The final expression for the action is
\begin{equation}
S_{\operatorname{total}}=S+S_{\operatorname{sources}}+S_{\operatorname{reg}}+S_{\operatorname{gf}}
\end{equation}
and the generating functional can be written in the form:
\begin{equation}
\begin{aligned}
&Z[J, j, \tilde{j}]=\int DV D\phi D\tilde{\phi} \, (\det(PV, M))^{N_{f}} \exp(iS+iS_{\operatorname{reg}}+iS_{\operatorname{sources}}+iS_{\operatorname{gf}}) ,
\end{aligned}
\end{equation}
where $J, j, \tilde{j}$ are the sources.
From this expression, one can obtain the propagators of the theory for the ordinary superfields and for the Pauli-Villars (massive) superfields:
\begin{equation}\label{prop1}
\begin{aligned}
&P\left(\phi_{\alpha, x}, \phi_{\beta, y}^{*}\right)=P\left(\tilde{\phi}_{\alpha, x}, \tilde{\phi}_{\beta, y}^{*}\right)=\delta_{\alpha \beta} \frac{\bar{D}_{x}^{2} D_{y}^{2}}{4 \partial^{2}} \delta_{x y}^{8} ; \\
&P\left(\Phi_{x}, \widetilde{\Phi}_{y}\right)=\frac{M \bar{D}^{2}}{\partial^{2}+M^{2}} \delta_{x y}^{8}; \\
&P\left(\Phi_{x}^{*}, \tilde{\Phi}_{y}^{*}\right)=\frac{M D^{2}}{\partial^{2}+M^{2}} \delta_{x y}^{8} ; \\
& P\left(\Phi_{x}, \Phi_{y}^{*}\right)=P\left(\tilde{\Phi}_{x}, \widetilde{\Phi}_{y}^{*}\right)=\frac{\bar{D}_{x}^{2} D_{y}^{2}}{4\left(\partial^{2}+M^{2}\right)} \delta_{x y}^{8}.
\end{aligned}
\end{equation}
The gauge field propagator reads as
\begin{equation}\label{prop2}
\begin{aligned}
&P\left(V_{x}, V_{y}\right)=2 e_{0}^{2}\left[-\frac{1}{R \partial^{2}}+\frac{1}{16 \partial^{4}}\left(\frac{\xi_{0}}{K}-\frac{1}{R}\right) \Big(\bar{D}^{2} D^{2}+D^{2} \bar{D}^{2}\Big)\right] \delta_{x y}^{8}.
\end{aligned}
\end{equation}
The vertices in this case can be easily constructed by the direct expansion of the action.
The regulator function $\frac{1}{R}$ will be called K4 in the following example. In this case, we omit the flavor indices; in the program, the powers of $N_{f}$ is determined by the number of closed matter loops.

The program is currently a console application, the data is read from a text file. At the input of the program, one must send
\begin{enumerate}
\item Calculation type, that is kind of contribution to effective action do we consider.
\item Various options, for instance, marking that we are working with supersymmetric quantum electrodynamics with $N_{f}$ flavors.
\item Expansion order and a number of loops.
\item List of propagators and vertices of the theory
\end{enumerate}
 Let us consider an example input file for the one loop calculation in the Feynman gauge (this means that $\xi_{0}=1$ and $K=R$ in (\ref{prop2})).
\begin{verbatim}
Type:
F#_1^1{-1}*F_1^1{1}
Option:
SQED
Nf^0
Loops:
1
Order:
2
Propagators:
V_1^1{1}*V_2^1{1}
2*i*e^0*I{1^-2}*K4{1}*d_{1}{12}
F#_1^1{1}*F_2^1{1}
-1/4*i*e^0*I{1^-2}*D_1(D#_2(d_{1}{12}))
Vertexes:
1/2*i*e*F#_1^1{6}*V_1^1{-3}*F_1_1{-2}
\end{verbatim}
In our example we consider a contribution to the effective action with two external matter superfields and examine the case of zero external momenta. We work in theory with $N_{f}$ flavors and calculate only the contributions that do not contain $N_{f}$ .

Now here is the output for this example:
\begin{verbatim}

File read correctly
1-loop diagrams are being generated
Completed100%
Diagrams created
Aligning momenta ...
Diagram generation completed
Diagram generation took 0.027 sec
Diagrams are being calculated...
0
-1/8*e^2*F#_1^1{-1}*F_2_2{1}*d_{-2}{12}*
D_2(D#_1(d_{2}{12}))*K4{2}*I{2^-2}*I{2^-2}
Diagram calculation completed
Diagram calculation took 0.018 sec
Collecting terms...
Collecting terms completed
Result:
-1/2*e^2*F#_1^1{-1}*F_1_2{1}*K4{2}*I{2^-4}
Total running time 0.052 sec
\end{verbatim}
Each step is explained, the generation process displays the percentage of what has been done so far. It then displays the time it took to generate. When diagrams are being calculated, for each graph its number (starting from 0) and an expression for it are displayed. At the end, the result is displayed. In this case, this expression is translated into analytical language as follows:
\begin{equation}
-\frac{1}{2} \int \frac{d^{4} k}{(2 \pi)^{4}} d^{4} \theta \, \phi_{\alpha}^{* }(0, \theta) \phi_{\alpha}(0, \theta) \frac{1}{R_{k}k^4}.
\end{equation}
In general, all input expressions are assumed to be integrated over the full superspace, and all output expressions over $d^4 \theta$ and over all loop momenta. External momenta in this case are set equal to zero, but if this is not the case, then integration is assumed over them as well. Since the connection of the effective action with the two-point Green's function is expressed by the equation
\begin{equation}
\begin{aligned}
&\Gamma^{(2)}_{\phi}= \frac{1}{4} \sum_{\alpha=1}^{N_{f}} \int \frac{d^{4} q}{(2 \pi)^{4}} d^{4} \theta\left(\phi_{\alpha}^{*}(-q, \theta) \phi_{\alpha}(q, \theta) +\widetilde{\phi}_{\alpha}^{*}(-q, \theta) \widetilde{\phi}_{\alpha}(q, \theta)\right) G\left(\alpha_{0}, \Lambda / q\right),
\end{aligned}
\end{equation}
then the contribution to the Green function (in which the external momentum is set equal to zero) has the form:
\begin{equation}
\left.\Delta G\right|_{q=0}=-2 e_{0}^{2} \int \frac{d^{4} K}{(2 \pi)^{4}} \frac{1}{K^{4} R_{K}}.
\end{equation}
This result agrees with the results of Ref. \cite{Kataev:2014gxa}. Moreover, it corresponds to formula (56) from \cite{Aleshin:2020gec} with $\xi_{0}=1$. In addition, it is worth noting that the program outputs an expression in the form of Euclidean integrals if all input expressions are also written in terms of Euclidean momenta. Changing the measure as a result of Wick's rotation and other transformations are done automatically by the program. 

\section{Technical information and timing}
\hspace*{\parindent}\label{Section_MSSM_NSVZ32323}
The program was used in calculations in up to three loops. The results of the three-loop calculation can be found in Ref. \cite{Shirokov:2022jyd}.

In the end, we present the time that the various calculations of this program took. The calculations were carried out on the following configuration:
 
\vspace{1em}
{\em
Operating system: Windows 10 x64 \\
Processor: AMD Ryzen 5 1600 Six-Core Processor 3.20 GHz \\
RAM: 8 GB \\
Compiler: GNU GCC Compiler\\
Compilation options: -march=native, -O3
}
\vspace{1em}

We will divide them into three parts depending on the degree of the number of flavors. Moreover, since this degree is determined by the number of closed matter loops, then some contributions exist only in a certain order. In addition, we will also consider the cases of the minimal ($\xi_0=1$) and non-minimal ($\xi_0\neq 1$) gauges, as well as the absence or presence of massive fields in the theory ($m$) . The timing is given in table. \ref{tab:Rv1},\ref{tab:Rv2},\ref{tab:Rv3}.
\begin{table}[h!]
\begin{center}
\begin{tabular}{|c|c|c|c|}
\hline
& 1 loop & 2 loops & 3 loops\\
\hline
$\xi_0=1$ & 0.052 sec & 0.14 sec & 2.6 sec\\
\hline
$\xi_0\neq 1$ & 0.067 sec & 0.57sec & 2 min 27 sec \\
\hline
\end{tabular}
\caption{\label{tab:Rv1}Timing, contribution $\sim N_{f}^0$}
\end{center}
\end{table}

\begin{table}[h!]
\begin{center}
\begin{tabular}{|c|c|c|c|}
\hline
& 2 loops & 3 loops\\
\hline
$\xi_0=1,\, m=0$ & 0.16 sec & 6.6 sec\\
\hline
$\xi_0\neq 1,\, m=0$ & 0.52 sec & 13 min 49 sec \\
\hline
$\xi_0=1, \, m\neq 0$ & 0.41 sec & 41.5 sec\\
\hline
$\xi_0 \neq 1, \, m \neq 0$ & 1.23 sec & 3 h 54 min\\
\hline
\end{tabular}
\caption{\label{tab:Rv2}Timing, contribution $\sim N_{f}^1$}
\end{center}
\end{table}

\begin{table}[h!]
\begin{center}
\begin{tabular}{|c|c|c|c|}
\hline
& 3 loops\\
\hline
$\xi_0=1,\, m=0$ & 4.2 sec \\
\hline
$\xi_0\neq 1,\, m=0$ & 6.6 sec \\
\hline
$\xi_0=1, \, m\neq 0$ & 35 sec \\
\hline
$\xi_0\neq 1, \, m \neq 0$ & 2 min 58 sec \\
\hline
\end{tabular}
\caption{\label{tab:Rv3}Timing, contribution $\sim N_{f}^2$}
\end{center}
\end{table} 
In the case of Feynman gauge ($\xi_0=1$), we considered vertices in which the number of external lines of the gauge field does not exceed the number of loops, diagrams where this is not the case are zero in this gauge according to (\ref{r3}). This slightly reduced the running time.

\section{Conclusion}

A program was created to generate and calculate the superspace part of Feynman diagrams for two-point Green functions of supersymmetric electrodynamics. It has been tested on calculations in one and two loops and used for calculations in three loops, including the ones in the non-minimal gauge ($\xi_{0}\neq 1$). The running time in most calculations turned out to be acceptable (usually no more than a few minutes), which makes it possible to further upgrade it for calculations in more interesting theories, such as the supersymmetric Yang-Mills theory.

\section*{Acknowledgements}

The author expresses deep gratitude to K.V. Stepanyantz for his help in studying methods of computing in superspace, as well as for careful reading of the text of the article and important comments. In addition, the author thanks S.V. Morozov for his help and valuable advices on the C++ language capabilities and the OpenMP package, as well as for careful reading of the article and important comments. The author also thanks A.L. Kataev for his valuable comments.

\end{document}